# THIRD HARMONIC GENERATION AND PHOTOLUMINESCENCE MEASUREMENTS IN ZINC OXIDE AND ALUMINUM DOPED ZINC OXIDE THIN FILMS GROWN BY ATOMIC LAYER DEPOSITION

Calford Odhiambo Otieno[1]

1. *Department of Physics, Kisii University P.O Box 408-40200, Kisii, Kenya.*

**Abstract**

Zinc Oxide is a thoroughly studied wide-bandgap semiconductor possessing excellent optical and electronic properties at room temperature. The renewed interest in this material has been generated by doping with various impurities in order to further enhance versatile optoelectronic responses for practical applications. Specifically, Aluminum-doped Zinc Oxide is an emerging transparent conducting oxide for photovoltaic applications. Here I propose to conduct a series of experimental studies on broadband optical nonlinearity as well as photoluminescence from Aluminum doped Zinc Oxide as a function of Aluminum doping. The results for this study includes studies include 1) Bandgap measurements 2) wavelength-dependent third harmonic generation and 3) one-photon-induced Photoluminescence. The most notable result is a multifold enhanced third-order nonlinear optical response from weakly doped Aluminum doped Zinc Oxide (up to 4%) in comparison with the undoped counterpart. The observed nonlinear optical trend as a function of Aluminum doping is correlated to the modification of the corresponding band structure. The Aluminum doping effect is extensively investigated in the context of absorption and Photoluminescence.

**I. Background**

Zinc Oxide is a very promising material for semiconductor device applications. It has a direct and wide bandgap in the near-ultraviolet (near-UV) spectral region ($E_g$ = 3.3 eV). An exciton in Zinc Oxide has a large binding energy of 60 meV so that excitonic emission can occur at room temperature [1-4]. Zinc Oxide crystallizes in the wurtzite structure (Fig.1a), and is available as large bulk single crystals. Its material properties have been extensively studied [5-8]. The use of Zinc Oxide as an active component in electronic devices has been hindered by the lack of control over its electrical conductivity [9-11]. However, significant improvement in the quality of Zinc Oxide single crystal substrates and fabrications of epitaxial films has led to numerous applications of Zinc Oxide in many devices including sensors, transistors, and light-emitting diodes [12-14]. Many Density functional calculation studies oxygen vacancies and native point defects in Zinc Oxide have also been performed and documented [15-18]. Bandgap engineering of Zinc Oxide can be achieved by alloying with Al, Mg, or Cd. For example, it has been observed that adding Aluminum to Zinc Oxide increases the bandgap, whereas adding Cd decreases the bandgap [19, 20]. Upon addition of impurities the alloys assume the wurtzite structure of the parent compound, causing a significant bandgap variation that is good for practical applications. Zinc Oxide can adopt a Zinc blende structure, but the wurtzite structure is more stable and common in the ambient condition due to the nature of its ionic bond [$Zn^{2+}O^{2-}$] (Fig. 1). The electronic band structure of the wurtzite Zinc Oxide has been studied both theoretically using first principle calculations and experimentally. All results indicate that ZnO is a direct-bandgap material where the conduction band minimum and the valence band maxima occurs at the center of Brillouin zone[21].

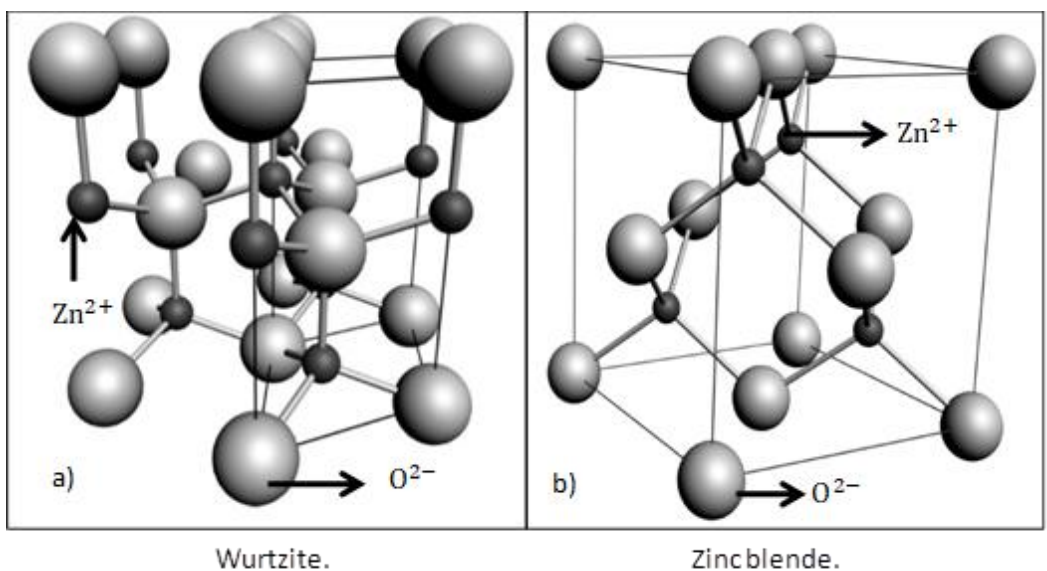

Fig 1: (a) Hexagonal wurtzite structure of ZnO with oxygen atoms shown as large spheres and zinc atoms small black spheres. (b) Zincblende structure of ZnO [21].

## II. Nonlinear Optical Theory

The theory of nonlinear optics follows from light-matter interaction, where the optical properties of a material are modified in the presence of an intense electromagnetic perturbation that in turn changes the light field. Nonlinear Optical interactions consist of two successive processes: First, an intense light beam (laser) induces nonlinear polarization in the medium. Second, the medium reacts on the light via time-varying nonlinear polarization, which is related to the acceleration of charges in the medium. The first process is governed by the constitutive equations, which relate the external field and induced polarization in terms of an optical susceptibility. The second process is governed by Maxwell's equations, which describe the generation of new frequencies in the presence of nonlinear polarization and governed by the nonlinear wave equation. Macroscopic polarization is the number of dipole moment per unit volume. Many Nonlinear optical processes of Zinc Oxide and Aluminum doped Zinc Oxide are so far documented as follows; Two-photon absorption (2PA) coefficient of bulk ZnO is $\beta = 5$ cm/GW at 532 nm [25]., The three-photon absorption (3PA) coefficient $\gamma = 5.4 \times 10^{-3}$ cm$^3$/GW$^2$ at 900 nm characterized via Z-scans and reported [26]. However, the relevant nonlinear optical dispersions, i.e., wavelength-dependences are not fully studied especially at longer wavelengths, far below the bandgap. Moreover, there are only a few reports on the nonlinear optical properties of aluminum doped Zinc Oxide at the single wavelength of 1064 nm [27, 28].
.

**IV. Optical Transmittance and Bandgap Measurements**

This section reports optical characterization of Zinc Oxide and Aluminum doped Zinc Oxide films grown by atomic layer deposition (ALD). Figure 2 shows typical optical transmission spectra from 250-nm-thick Aluminum doped Zinc Oxide samples with various Al-doping rates characterized by a UV-VIS spectrometer. Details on sample preparation is detailed and reported [35]

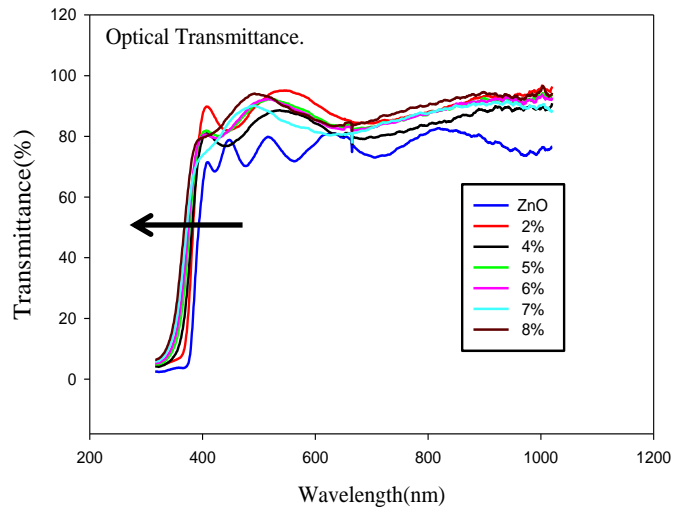

Fig. 2: Optical transmittance spectra of Zinc Oxide (blue trace) and Aluminum doped Zinc Oxide deposited on glass substrates with various Al-doping rates ranging from 2% to 8%. The absorption edge is blue shifted with increased Al-doping.

As the Al-doping rate increases from 2% to 8%, a noticeable blue shift in the absorption edge occurs as indicated by the arrow. Figure 4 show that all Aluminum doped Zinc Oxide films exhibit about 85% transmittance in the visible and near-IR regions. The oscillatory behaviors result from interference effects between the film and the substrate. Absorbance A is related to the sample thickness d and transmittance T and given by

$$T = A \exp(-\alpha d), \qquad (9)$$

where $\alpha$ is the one-photon absorption coefficient assuming negligible reflection. The optical bandgap is determined using the Tauc's model [35]. In a direct-gap semiconductor the absorption coefficient and the bandgap energy $E_g$ are related by

$$(\alpha h\nu) = C(h\nu - E_g)^n, \qquad (10)$$

where the exponent is $n = 0.5$ and $h\nu$ is the photon energy [36].

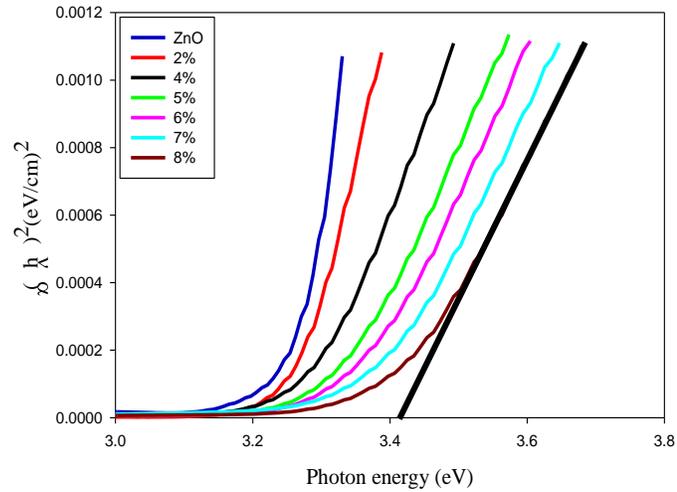

Fig. 3: Plot of $(\alpha h\nu)^2$ as a function of photon energy for Zinc Oxide and Aluminum doped Zinc Oxide, where the extrapolation to the energy axis corresponds to the optical gap for 8% Aluminum doped Zinc Oxide.

The optical bandgap was obtained from a graph of $(\alpha h\nu)^2$ versus photon energy, as shown in Fig. 3. Experimental bandgap seems to be partially affected by the possible formation of impurity phases at high Al doping as evidenced by a gradual increase in $(\alpha h\nu)^2$ across the bandgap, the so-called Urbach tail [37]. By extrapolating the linear part of the curves to the photon energy axis, each optical gap was determined. The measured optical gap due to doping ranged from 3.25 eV for the as-grown Zinc Oxide to 3.42 eV for the 8% Aluminum doped Zinc Oxide. The blue shift in the absorption edge is due to change in the Fermi level towards the conduction band leading to the bandgap widening. This bandgap widening occurs due to blocking of low-energy transition to the conduction band and is known as the Burstein-Moss effect [38, 39]. The energy shift is related to the carrier concentration by $\Delta E_g =$

$\frac{h^2}{8m^*}\left(\frac{3n_e}{\pi}\right)^{\frac{2}{3}}$ where h is the Planck's constant, $m^*$ is the reduced mass of the associated bands, and $n_e$ is the carrier density [40, 41]. Therefore, the optical gap energy is given by

$$hf_{min} = E_g + \frac{h^2 k_F^2}{2}\left(\frac{1}{m_c} + \frac{1}{m_v}\right) = E_g + \frac{h^2}{8m^*}\left(\frac{3n_e}{\pi}\right)^{\frac{2}{3}}, \qquad (11)$$

where $k_F$ is the Fermi wave vector, and $m_c$ and $m_v$ are the effective masses of the conduction band and valence band, respectively. The measured gap as a function of Al-doping rate is shown in Fig. 4. The dots represent experimental values and the solid curve is the Burstein-Moss effect incorporated with a non-parabolic conduction band model [42]. The observed trend is similar to the trends reported in Ref. [41].

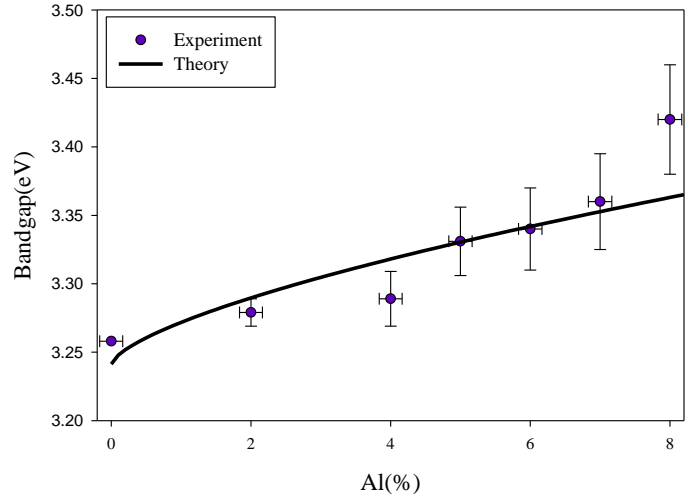

Fig. 4: Experimental bandgap as a function of Al-doping (dots), superimposed by Eq. (11) assuming a non-parabolic conduction band model.

## III. Experimental Details On Nonlinear Third Harmonic Generation Measurements

Experiments using a solid-state (PL 2250) series Nd:YAG laser operating at a near-IR wavelength (1.064 μm). This fundamental beam is fed into an H400 harmonic unit where its frequency is doubled and tripled by SHG and cascade process (SFG of SHG and fundamental beams) in appropriate nonlinear crystals. The frequency-tripled UV (355 nm) and the fundamental IR (1.064 μm) are collinearly fed into a PG403 optical parametric oscillator (OPO). These successive nonlinear processes allow me to tune the output wavelength from 210 nm to 4.450 μm. The output pulse width is about 30 ps with a repetition rate of 50 Hz. The laser pulse from the optical parametric Oscillator is focused on Aluminum doped Zinc Oxide using a positive lens in which the sample position Z can be continuously varied near the Gaussian beam waist via Z-scan. The nonlinear optical effects and and Photoluminescence from the samples are collected via a collection lens using transmission or reflection geometry by a fiber-optic bundle, which is coupled to a selective-grating (1800, 600, and 300 grooves/mm) spectrometer equipped with a charge coupled device (CCD) camera (Synapse) as well as an extended InGaAs (Symphony) detector. Broadband nonlinear optical data are generated by calibrating the wavelength-dependent optical and quantum efficiencies of optical components and detectors in the collection system. Exciton recombination lifetimes are measured via a C10627 Hamamatsu Streak Scope with a 5 ps resolution coupled to a C9300 Hamamatsu Digital CCD Camera.

## V. Results for Broadband Third Harmonic Nonlinear Optical Measurements

Presented results indicate Zinc Oxide exhibited strong THG when the fundamental wavelength was varied between 1.2 μm and 2.1 μm. The corresponding THG wavelengths range from 400 nm to 700 nm. Since the glass substrate as well as other optical components can generate THG, careful measurements and data processing were required to eliminate or minimize these background THG signals. First, I inserted a short-pass filter (~50% transmittance in the visible region) just before the fiber-optic cable in order to suppress the remnant fundamental beam that can cause additional THG at the fiber input. Second, THG Z-scan to locate an optimized sample position to maximize the THG counts from Zinc Oxide with minimal THG from the glass substrate, where the latter was determined by another THG Z-scan on a bare glass substrate with the same thickness. Finally, background THG signals from the substrate were measured at each wavelength and subtracted in the THG data accordingly to single out THG counts from Zinc Oxide only. A similar procedure was employed to THG measurements of AZO samples.

The blue traces in Fig. 5 show the wavelength-dependent THG spectra from Zinc Oxide when the fundamental wavelength was varied in the experimental range. The black traces in Fig. 6 correspond to those from 4%-doped Aluminum doped Zinc Oxide under the same experimental conditions. Significant decrease in the THG responses was observed when the THG wavelength approaches the optical bandgap ~3.22 eV for Zinc Oxide and ~3.32 eV for 4%-doped Aluminum doped Zinc Oxide. This is as a result of efficient bandgap absorption of the produced THG beam. However, at higher wavelengths, strong THG was observed. Note that the THG response is enhanced by more than an order of magnitude in Aluminum doped Zinc Oxide when compared with Zinc Oxide. One of the goals is to find the best Aluminum doping level to maximize the THG response. For the film samples the absolute value of $\chi^{(3)}$ can be estimated from the relation

$$I(3\omega, d) = \frac{9\omega^2|\chi^{(3)}|^2 I^2}{16 n_{3\omega} n_\omega^3 c^4 \epsilon_0^2} \text{sinc}^2\left(\frac{\Delta k d}{2}\right) I^3(\omega), \quad (12)$$

where d is the interaction length, i.e., the sample thickness, c is the speed of light in vacuum, $I(\omega)$ and $I(3\omega, d)$ are the fundamental and THG intensities, and $n_\omega$ and $n_{3\omega}$ are the refractive indices at fundamental and THG wavelengths, respectively. The THG intensity depends on the phase-matching factor, $\Delta k = \frac{6\pi}{\lambda}[n_{3\omega} - n_\omega]$ for THG. THG is typically non-phase-machable in condensed matter systems because of a large phase-mismatch. Considering a submicron thickness of the samples (d = 250 nm).

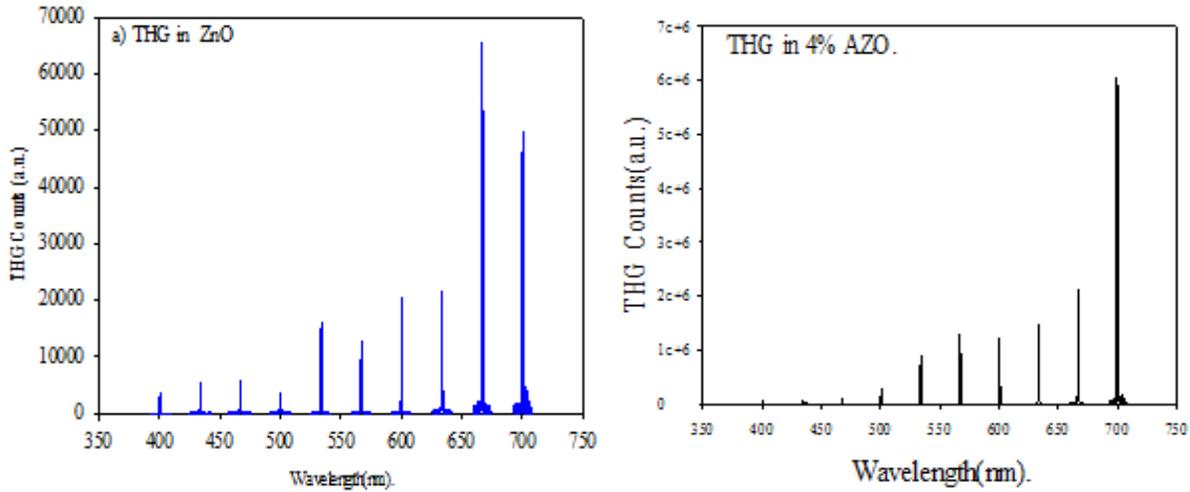

Fig.5: Observed broadband THG spectra from ZnO. The THG counts plummet as the wavelength approaches the bandgap, indicating strong bandgap absorption. Wavelength-dependent THG response from 4%-doped AZO, showing enhanced THG counts with Al-doping.

The relative THG counts for ZnO and two doping rates of ZnO and 4% are compared in Fig. 9. It is evident that THG markedly increases with Al-doping. I will further examine THG for higher doping rates over a broader wavelength range. Of particular interest is THG response at mid-IR in which oxygen-induced fundamental absorption can significantly suppress the THG efficiency. I will examine this effect to identify the practical working range for NLO applications involving third-order NLO processes.

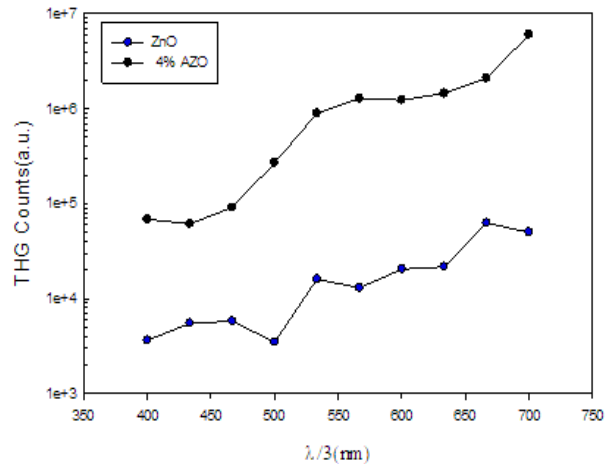

Fig. 6: Broadband THG counts on a semi-log scale for ZnO (blue) and 4%-doped (black) AZO.

## III. One –Photon-Induced room temperature Photoluminescence

It is generally agreed that ZnO has three major emission peaks. The UV located at ~380 nm is attributed to free exciton emission, the green peak at 510 nm is due to oxygen vacancies and a red peak at 615 nm is attributed to oxygen interstitials [48]. I conducted room-temperature PL experiments on undoped ZnO film using the 355 nm excitation source. Figure 8 illustrates typical time-integrated PL spectra when the excitation energy was varied in the range of 0.8 µJ − 10.25 µJ. At low excitation levels the PL spectra are broad and asymmetric this is the X-band emission and is attribute to the radiative recombination of excitons directly or indirectly via phonons. Upon increasing excitation energy a sharp peak appears and builds up superlinearly, indicating stimulated emission. This stimulated emission is as a result of scattering of two thermal excitons causing radiative recombination of the photon-like excitons near the polariton bottleneck [49]. Unlike typical ZnO samples no observation of any random lasing [50,51], indicative of a high sample quality.

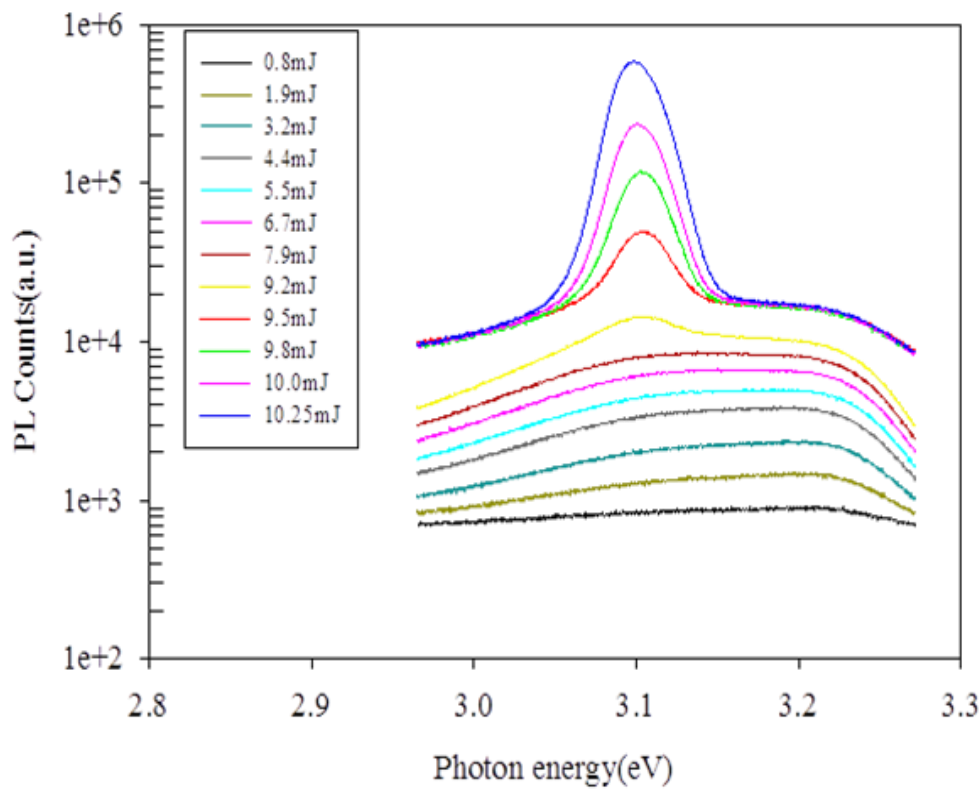

Fig.8: Time-integrated PL spectra of ZnO at room temperature under various excitation levels, showing broad exciton peaks at low excitation and stimulated emission at high excitation.

## V. CONLUSIONS

The results for nonlinear third harmonic generation is presented and discussed, the most notable result is a multifold enhanced third-order nonlinear optical response from weakly doped Aluminum doped Zinc Oxide (up to 4%) in comparison with the undoped counterpart attributed to the electronic nonlinear response. The photoluminescence spectra are broad and asymmetric at low excitation and are attributed to the radiative recombination of excitons directly or indirectly via phonons. Upon increasing excitation energy a sharp peak appears and builds up superlinearly, indicating stimulated emission. This stimulated emission is as a result of scattering of two thermal excitons causing radiative recombination of the photon-like excitons near the polariton bottleneck.